\documentclass {jetpl}
\twocolumn

\lat

\title {Nature of the electronic states involved in the chemical bonding and 
superconductivity at high pressure in SnO}

\rtitle {Nature of the electronic states involved in the chemical bonding and 
superconductivity at high pressure in SnO}

\sodtitle {Nature of the electronic states involved in the chemical bonding and 
superconductivity at high pressure in SnO}

\author {J.\,A.\,McLeod$^+$\thanks {e-mail: john.mcleod@usask.ca}, 
A.\,V.\,Lukoyanov$^{*,**}$, E.\,Z.\,Kurmaev$^{*}$, L.\,D.\,Finkelstein$^*$, A.\,Moewes$^+$}

\rauthor {J.\,A.\,McLeod, A.\,V.\,Lukoyanov, E.\,Z.\,Kurmaev et al.}

\sodauthor {J.\,A.\,McLeod, A.\,V.\,Lukoyanov, E.\,Z.\,Kurmaev, L.\,D.\,Finkelstein, A.\,Moewes}

\address {$^+$Department of Physics and Engineering Physics, University of Saskatchewan, 
116 Science Place, Saskatoon, Saskatchewan S7N 5E2, Canada\\~\\
$^*$Institute of Metal Physics, Russian Academy of Sciences--Ural Division, 
620990 Yekaterinburg, Russia\\~\\
$^{**}$Ural Federal University, 620002 Yekaterinburg, Russia}

\dates {\today}{*}

\abstract {We have investigated the electronic structure and the Fermi surface 
of SnO using density functional theory (DFT) 
calculations within recently proposed exchange-correlation potential (PBE+mBJ) 
at ambient conditions and high pressures up to 19.3 GPa where superconductivity 
was observed. It was found that the Sn valence states (\textit{5s}, \textit{5p}, 
and \textit{5d}) are strongly hybridized with the O \textit{2p} states, 
and that our DFT calculations are in good agreement with O \textit{K}-edge X-ray 
spectroscopy measurements for both occupied and empty states. 
It was demonstrated that the metallic states appearing under pressure 
in the semiconducting gap stem due to the transformation of the weakly 
hybridized O \textit{2p} -- Sn \textit{5sp} subband corresponding 
to the lowest valence state of Sn in SnO.
We discuss the nature of the electronic states 
involved in chemical bonding and formation of the hole and electron pockets 
with nesting as a possible way to superconductivity.}

\begin {document}
\maketitle

Guided by research in iron pnictides, superconductivity was observed recently 
in the layered ($\alpha$-PbO-type) crystal SnO at high pressure with T$_c$ = 1.4 K 
and B$_{c2}$ = 0.6 T at P = 9.3 GPa~\cite{Forthaus10}. 
It was found that electronic structure of SnO at 7 GPa is that of a metal 
with a hole pocket at the $\Gamma$-point and electron pockets at the M-point 
of Brillouin zone, i.e. both the crystal and electronic structure are similar 
to those of superconducting iron pnictides (FeAs and FeSe-based compounds)
~\cite{Forthaus10}. 
On the other hand, unlike the iron pnictides, SnO is nonmagnetic and does not 
show any structural phase transitions up to P = 19.3 GPa~\cite{Wang04}. 

Thus far, the available electronic structure calculations of SnO at high pressure 
have been performed for high pressures up to 10 GPa (where V/V$_0$=0.90) and 
have been limited to calculation of the E(k) dispersion 
curves~\cite{Forthaus10,Christensen05,Christensen06,Lefebvre98}. 
The minor role of Sn \textit{5p} states near the Fermi level at ambient pressure 
was examined by Watson \cite{Watson01}, he showed that these states are shifted 
to lower energies due to the hybridized combination of the O \textit{2p} 
and Sn \textit{5s} states just below the Fermi level to provide stability 
for the litharge structure.  

The superconductivity of SnO and layered iron pnictides is increasingly being 
examined from the perspective of the Fermi surface which is often calculated 
by density functional theory (DFT). Because the superconductivity of SnO arises 
during a semiconductor to metal transition, the nature of the band gap closure 
is important. Since DFT routinely underestimates the magnitude of the band gap 
\cite{Dufek94} it is important to compare the calculated electronic structure 
to experimental measurements.

In the present paper we have calculated the electronic structure for a range 
of pressures from ambient to 19.3 GPa using a variety of exchange-correlation 
functionals, and in particular analyzed 
the electronic structure and Fermi surface for SnO at P = 0.0, 5.1 (the onset 
of superconductivity \cite{Forthaus10}), 9.2 (the maximum T$_c$ \cite{Forthaus10}), 
and 19.3 (where superconductivity vanishes) GPa. 

The calculations on SnO at ambient pressure are compared to non-resonant oxygen 
\textit{K}-edge X-ray emission spectroscopy (XES) measurements, performed 
at Beamline 8.0.1 of the Advanced Light Source (ALS) at Lawrence Berkeley National 
Laboratory \cite{Jia95}, and oxygen \textit{K}-edge X-ray absorption spectroscopy 
(XAS) measurements, performed at the spherical grating monochromator beamline 
of the Canadian Light Source (CLS) at the University of Saskatchewan~\cite{Regier07}. 
The absorption measurements were acquired in the bulk sensitive total fluorescence 
yield (TFY) mode \cite{Bou03,Kang05}. Commercially available SnO powder (Alfa Aesar, 99\% purity) 
was used for the spectroscopy measurements. The powder was pressed onto clean 
indium foil and measured without further preparation.

Our calculations were performed using the WIEN2K code \cite{Blaha01} which 
is based on the full-potential linear augmented plane-wave plus local-orbital 
(FP-LAPW+lo) method with scalar-relativistic corrections. 
The spin-orbit coupling was not accounted for. For our exchange-correlation 
functional we have used the local density approximation (LDA) and the Perdew-Burke-Ernzerhof 
(PBE) variant of the generalized gradient approximation~\cite{Perdew96}. 
To improve the calculated band gaps, we have also added a calculation cycle 
to the PBE calculation using the modified Becke Johnson (mBJ) exchange potential~\cite{Tran09}. 
We set the cut-off parameter $R^{min}_{MT}K^{max}$=7 (the product of the smallest 
of the atomic sphere radii $R_{MT}$ and the maximum plane-wave wavenumber $K$) 
for the expansion of the basis set. The calculation was performed on 
a 14$\times$14$\times$11 or a 14$\times$14$\times$12 k-point grid 
for pressures below and above 9 GPa, respectively. The experimental crystal 
structure of SnO under pressure was used~\cite{Wang04}. The same sphere radii 
were used for both tin and oxygen atoms, and chosen such that the spheres 
were nearly touching. 

\begin {figure}
\includegraphics[width=3in]{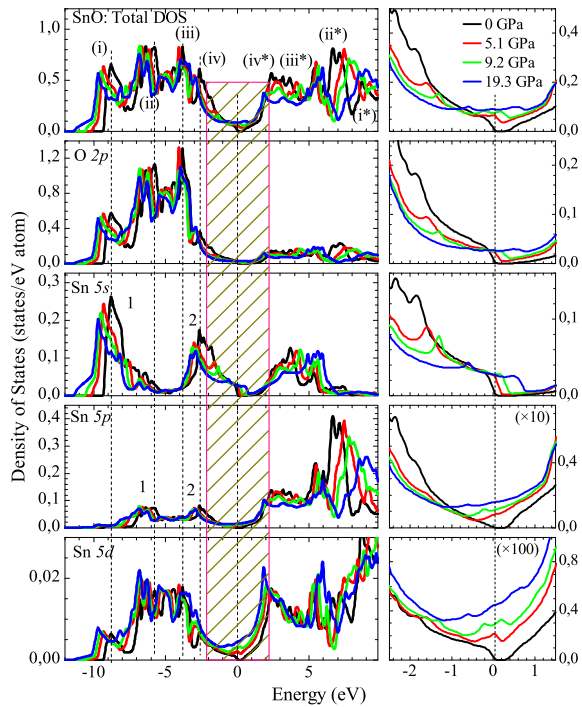}
\caption {Fig. 1. Total and partial densities of states of SnO for ambient 
pressure and pressures 5.1, 9.2, and 19.3 GPa. The key bonding (i - iv) and anti-bonding (iv* - i*) 
features are labeled and discussed in the text. The PBE + mBJ calculation is shown here, apart from 
the band gap the other calculations give very similar DOS spectra.}
\label{fig1}
\end{figure}

The total and partial densities of states (DOS) of SnO at ambient and pressures 
of 5.1, 9.2, and 19.3 GPa are presented in Fig. \ref{fig1}. At normal conditions 
SnO is a semiconductor with a small indirect gap. The LDA calculation predicts 
a gap of 0.084 eV, the PBE calculation predicts a gap of 0.033 eV, 
and the PBE + mBJ calculation predicts a gap of 0.254 eV. As one would expect, 
the O \textit{2p} states dominate the valence band and mimic the shape 
of the total DOS. In the conduction band the main contribution near the bottom 
of the band is again due to O \textit{2p} hybridized with the Sn \textit{5s} 
and \textit{5p} states. Above 5 eV the Sn \textit{5p} 
states are comparable with the O \textit{2p} states, and the Sn \textit{5s} 
contribution becomes almost negligible. McLeod {\it et al.} \cite{McLeod10} 
demonstrated that O \textit{2p} electrons in oxides occupy donor cation states 
(in our case the Sn \textit{5s} states) and in addition to this what would 
normally be considered the nearest excited states (in our case 
the Sn \textit{5p} and \textit{5d} which would formally be considered 
nonbonding) are involved in chemical bonding.

In the total DOS of SnO, see Fig. \ref{fig1}, one can identify four 
bonding Sn-O subbands in the occupied part: i, ii, iii, 
and iv, and corresponding antibonding i*, ii*, iii*, and iv* subbands 
in the empty part. These bonding and antibonding subbands are almost 
equally energetically separated from the Fermi level that was used for 
their identification. The i, ii, and iii subbands are the main bonding 
subbands of the Sn-O ion-covalent interaction. They have 
the dominant O \textit{2p} over the cation contribution and concurrent 
O \textit{2p} maximum energy and energies of partial cation maximum, namely, 
the i subband O \textit{2p} DOS is located at the Sn \textit{5s}(1), 
the ii subband -- at the Sn \textit{5p}(1), the iii subband -- 
at the Sn \textit{5d}. Similar order of states: \textit{ns}, \textit{np}, 
and \textit{nd} was noted previously in the oxides of the elements 
of IIb group of the periodic table \cite{McLeod10}. 

The important peculiarity of the partial Sn \textit{5s} DOS and 
Sn \textit{5p} DOS is a separation into two parts 1 and 2, while 
the main bonding subbands i and ii are formed at the energies 
of Sn \textit{5s}(1) and Sn \textit{5p}(1) separated by approx. 2 eV. 
In contrast to them, Sn \textit{5s}(2) and Sn \textit{5p}(2) parts 
have the same energy; at this energy the iv weakly bonding 
subband is formed together with the relatively weak O \textit{2p} 
maximum of the DOS. The weak character of the Sn-O bonding 
in the iv subband is reflected in the comparable contributions 
of the O \textit{2p} and Sn \textit{5s,~5p}, while the O \textit{2p} 
states dominate in the i, ii, and iii subbands. In the simplified 
description of chemical bonding of SnO one would say that this iv subband 
corresponds to the unbonding electrons responsible for the lowest valence 
state of Sn, and the i, ii, and iii subbands correspond to 
the valent-active Sn \textit{5s,~p,~d} electrons making strong 
ion-covalent Sn-O interaction.

Applied pressure results in the approaching interacting Sn and O atoms, 
and all structural elements in the i-iv subbands are shifted down in energy, 
demonstrating the increase of bonding effect. The strongest bonding effect can be 
found for the deepest bonding subbands i and ii. Antibonding subbands iv*-i* 
under pressure are shifted to the higher energies demonstrating the growth 
of the antibonding effect. 

In the energies from --2 to 2.2 eV the tail parts of the iv and iv* subbands 
behave differently under pressure. In contrast to the shifts of the main maxima 
of the iv and iv* subbands, the tail parts are shifted toward each other closing 
the energy gap and increasing the number of states at the Fermi level, 
see Figs. \ref{fig1}, \ref{fig3}, \ref{fig4}c. The formation of the metallic states 
at the Fermi level is shown in detail in the right side of Fig. \ref{fig1}. 
Thus these metallic states originating from the weakly bonding 
iv and iv* subbands attribute to the lowest valency state of Sn.

\begin {figure}
\includegraphics[width=3in]{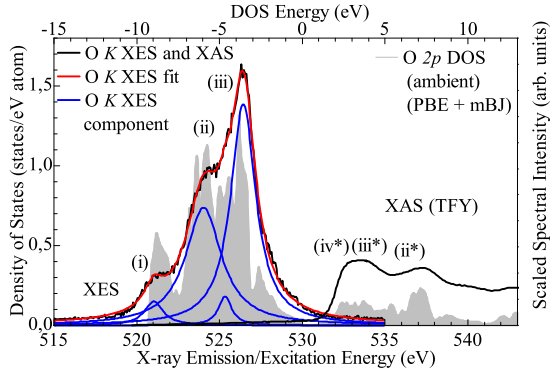}
\caption {Fig. 2. The measured O \textit{K}-edge XES and XAS. 
The XES spectrum has been fit with 4 pseudo-Voigt components, three 
of which agree with the previously discussed valence band intervals. 
The ambient O \textit{2p} DOS is also plotted, it has been aligned with 
the XES spectrum. The XAS spectrum shows the same two intervals as the calculated DOS.}
\label{fig2}
\end{figure}

Since the Sn states are involved in hybridization with the O \textit{2p} states, 
experimental spectroscopic investigations of the O \textit{2p} states can reveal 
significant information about all electronic states of SnO. 
The calculated O \textit{2p} DOS for SnO at ambient pressure is compared with 
measured XES (probing the valence states) and XAS (probing the conduction states) 
spectra in Fig. \ref{fig2}. 
The O \textit{K}-edge XES reproduces the calculated O \textit{2p} valence states 
accurately, and decomposing the measured spectrum into 4 pseudo-Voigt components 
reproduces three of the four valence intervals. The last interval (iv) is likely 
hidden in the experimentally broadened edge of the much larger iii interval. 
The O \textit{K}-edge XAS spectrum are also found in good agreement with 
the calculated spectrum.

%especially keeping in mind the influence of the O \textit{1s} 
%core-hole which locally moves conduction states to lower energies. 

\begin {figure}
\includegraphics[width=3in]{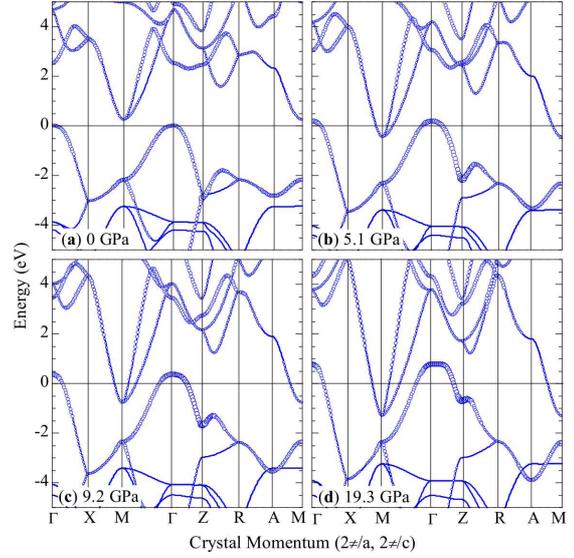}
\caption {Fig. 3. Band structure plots of SnO at ambient,  5.1 GPa, 9.2 GPa, 
and 19.3 GPa calculated with the PBE + mBJ exchange-correlation functional. 
The Sn \textit{5s} contribution to the band structure is proportional 
to the width of the line.}
\label{fig3}
\end{figure}

At ambient pressure we find the top of the valence band is formed by 
a dispersive band approaching the $\Gamma$-point, in agreement with 
the previously reported calculations~\cite{Forthaus10,Christensen05}. 
This band corresponds to the Sn \textit{5s} states hybridized with 
the O \textit{2p} states, as shown in the partial DOSes 
in Fig. \ref{fig1} and the Sn \textit{5s} contribution to the bands in Fig. \ref{fig3}. 
Increasing pressure results in the broadening of this band and 
causes the appearance of a hole pocket near the $\Gamma$ point. 
Increasing pressure also causes the appearance of electron pockets 
near the $M$ point due to the broadening of bands having a little 
Sn \textit{5s} contribution in comparison with the subband in the $\Gamma$ point. 
%Our band structure calculations show pressure-induced hole pockets 
%at the $\Gamma$ point and electron pockets at the $M$ point, 
%in agreement with previously reported calculations \cite{Christensen05,Forthaus10}.
The calculations for all exchange-correlation functionals predict 
the same general shape of the electron pocket at the Fermi level 
for SnO under high pressure, while the different exchange-correlation 
functionals predict slightly different sizes for the hole pocket.
The direct gap at the $\Gamma$ point is 2.1 eV for the LDA and PBE 
calculations, and 2.5 eV for the PBE + mBJ calculation. Early measurements 
of the optical gap in polycrystalline SnO 
suggest the gap should be between 2.5 and 3.0 eV \cite{Guerts84}. 

%The evolution of the electron and hole pockets is supported by the electronic structure: under applied pressure the predominant trend is for the bulk of all states in the first three valence intervals (i-iii) to be shifted to lower energies. 
%This shift is caused by both an increase in hybridization due to the decreasing volume, and a decreasing number of Sn \textit{5s} electrons (see Fig. \ref{fig4}) causing a reduced screening of the Sn core potential. 
%%Indeed, as shown in Fig. \ref{fig4}, under pressure the number of valence Sn \textit{5s} electrons decreases while the number of Sn \textit{5p} and \textit{5d} electrons grows slightly.

The correlation between the valence O \textit{2p} states and 
the Sn \textit{5s},\textit{5p},\textit{5d} states tells a more nuanced story. 
The correlation here is defined as $C_{a}(E) = \int \rho_{O \mathit{2p}}(E') \rho_a(E-E') dE'$ 
for partial DOS $\rho_i$ where $a$ = Sn \textit{5s}, \textit{5p}, or \textit{5d}.
This function determines the degree of similarity between two spectra, and the energy 
of the peak of this function indicates the overall shift between the two spectra. 
This energy shift is shown in Fig. \ref{fig4}b.
Note that while the Sn \textit{5s} states contribute to the hole pocket 
at the Fermi level, the bulk of the Sn \textit{5s} states are at lower 
energies than the O \textit{2p} states. Further, as pressure increases, 
the Sn \textit{5s} states shift to lower energies faster than the O \textit{2p} states. 
The \textit{5d} states, which contribute minimally to the valence band, 
almost exactly track the position of the O \textit{2p} states in energy. 
Most importantly is that PBE + mBJ calculation predicts that the Sn \textit{5p} 
states shift to higher energies than the O \textit{2p} states roughly concurrent 
with the onset of superconductivity (at 7 GPa, rather than 5.1 GPa). This is related 
to the increased influence of the electron pockets at the $M$ point.

\begin{figure}
\includegraphics[width=3in]{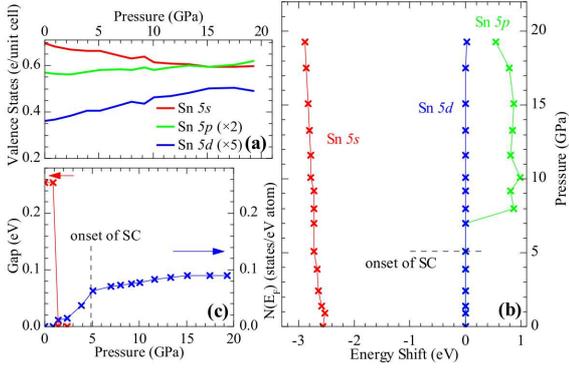}
\caption {Fig. 4. The number of valence states of Sn (\textbf{a}), 
the correlation shift between the Sn and O \textit{2p} states (\textbf{b}), 
and the band gap and number of states at the Fermi level (\textbf{c}).}
\label{fig4}
\end{figure}

Another change in the electronic structure concurrent with the onset of superconductivity 
is the saturation of the number of states at the Fermi level (N(E$_\mathrm{F}$)), shown in Fig. \ref{fig4}c. 
While the number of states at the Fermi level increases immediately after applied 
pressure has closed the band gap, N(E$_\mathrm{F}$) starts to saturate at 5.1 GPa in Fig. \ref{fig4}c. 
This is because the low pressure grown in N(E$_\mathrm{F}$) is mainly driven 
by increasing contributions of O \textit{2p} and Sn \textit{5s} states which 
reach a maximum at 5.1 GPa, while the Sn \textit{5p} states continue to increase 
their contribution this is only a minor addition to the total number of states 
(see the right side Fig. \ref{fig1}). 

\begin{figure}
\includegraphics[width=3in]{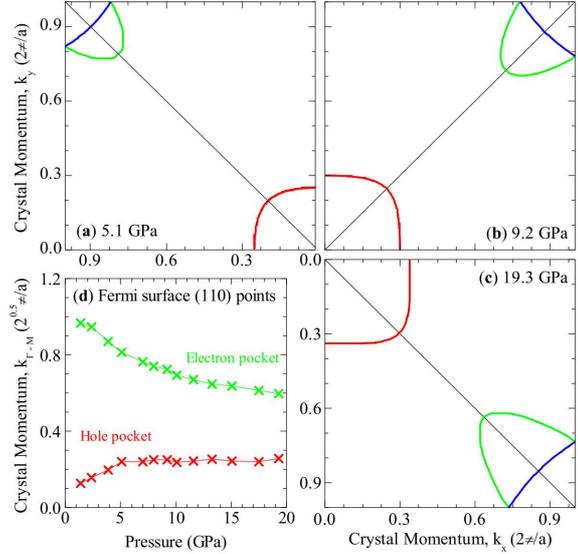}
\caption {Fig. 5. One quadrant of the Fermi surface in the $k_x$, $k_y$ plane 
(at $k_z = 0$) for SnO at 5.1 GPa (\textbf{a}), 9.2 GPa (\textbf{b}), 
and 19.3 GPa (\textbf{c}), calculated with the PBE + mBJ functional. 
Since our calculation suggests SnO at ambient pressures is an insulator, 
the Fermi surface in this case is trivial. 
In panel \textbf{d}, the nesting vector in the (110) direction from 
the corner of the $\Gamma$ point hole band to the opposite outer $M$ 
point electron band is also shown for various pressures.}
\label{fig5}
\end{figure}

\begin{figure}
\includegraphics[width=3in]{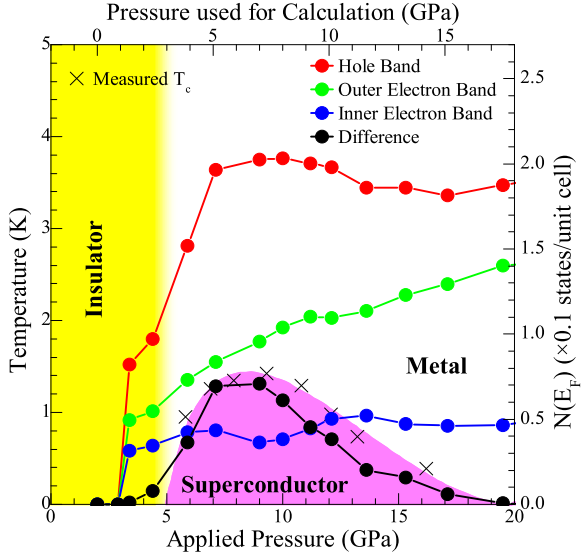}
\caption {Fig. 6. Number of states at the Fermi level of the bands 
forming electron and hole pockets and their difference (hole band) -- 
(inner and outer electron bands) vs. pressure. For comparison the phase diagram 
and pressure dependence of superconducting transition temperature from 
\cite{Forthaus10} are given accounting for the underestimated by \~1.5 GPa 
energy gap closure in the calculation.}
\label{fig6}
\end{figure}

As seen from the experiment and the calculated partial DOSes in Fig. \ref{fig1}, 
the states near the Fermi level mostly contain O \textit{2p} states, hybridized with 
the Sn \textit{5s} and \textit{5p} states. 
With increasing pressure these states are distorted into an increasing hole pocket at 
the $\Gamma$ point and two increasing electron pockets at the $M$ point, see Fig. \ref{fig5}. 
The distance between these two surfaces in the (110) direction decreases with increasing 
pressure, and the electron pockets grow faster than the hole pocket. 
In particular note that in Fig. \ref{fig3} the behaviour of the bands at the Fermi 
surface is essentially identical in the $\Gamma$ - X direction. 
%but the $\Gamma$ 
%- M direction shows some sensitivity to the choice of exchange-correlation functional. 
%As shown in Fig. \ref{fig5}, this sensitivity is especially marked 
%in the hole pocket for pressures below 5 GPa.
As one can see from Fig. \ref{fig6} showing the number of states at the Fermi level 
from electron and hole pockets, the hole carriers dominate over the whole region of 
superconductivity in SnO. The difference of the (hole band) -- (inner and outer electron bands) 
carriers tracks the behaviour of T$_c$ suggesting the hole-type of superconductivity 
in SnO at high pressure.  

In conclusion, we have investigated electronic structure and Fermi surface 
of SnO at ambient and high pressures to clarify the electronic states near 
the Fermi level involved in chemical bonding and superconductivity using 
\textit{ab initio} calculations. 
%We have found that the valence states of Sn (\textit{5s}, \textit{5p}, 
%and \textit{5d}) are strongly hybridized with O \textit{2p} states.
It was found that the semiconducting gap in SnO is closed under pressure 
by the weakly hybridized O \textit{2p} Sn \textit{5s,~p} states, located 
in the semiconductor SnO near the top of the valence band and the bottom 
of the conduction band. These states stabilize the lowest valence state 
of Sn but don't take part in strong ion-covalent O \textit{2p} -- 
Sn \textit{5s,~p,~d} interaction in the Sn-O layers. Under pressure 
they become more metallic and their tail parts fill the energy gap due 
to the increase of Sn-Sn interaction of the nearest layers.
It seems that compounds with cations in the lowest valency are perspective 
for new superconductors, and the marked similarity of SnO with FeSe and 
Fe-pnictide based superconductors is probably not accidental.
With the increasing pressure the size of hole and electron pockets 
at the $\Gamma$ and $M$ points, respectively, increases, leading to 
the nesting effect possibly involved in the formation of superconductivity. 

We acknowledge support of the Russian Foundation for Basic Research (Projects 11-02-00022, 
10-02-00046, and 10-02-00546), the Natural Sciences and Engineering 
Research Council of Canada (NSERC), the Canada Research Chair program, 
MK-3376.2011.2, the scientific program of the Russian Federal Agency of Science and 
Innovation under Project No. 02.740.11.0217, partial support of the scientific program 
``Development of Scientific potential of Universities'' under Project No. 2.1.1/779 
and the scientific program of the Russian Federal Agency of Science 
and Innovation under Project No. 02.740.11.0217.

\end{document}